\documentclass[showpacs,twocolumn,pre,floatfix]{revtex4}

\usepackage[dvips]{graphicx}
\usepackage[dvips]{color}
\newcommand{\href}[2]{#2}

\usepackage{amsmath}
\usepackage{amssymb}
\usepackage{bm}
\usepackage{times}

\newcommand{\vct}[1]{\mathbf{#1}}
\newcommand{\uvct}[1]{\mathbf{\hat{#1}}}

\newcommand{\grad}{\bm{\nabla}}

\begin{document}

\title{Colloid-colloid and colloid-wall interactions in driven suspensions}
\date{\today}

\author{Matthias Kr{\"u}ger}
\email{matthias.krueger@uni-konstanz.de}
\altaffiliation[Current address: ]{Fachbereich Physik, Universit{\"a}t
Konstanz, D-78457 Konstanz, Germany}
\affiliation{Max-Planck-Institut f{\"u}r
Metallforschung, Heisenbergstr.\ 3, 70569 Stuttgart, Germany, and \\
Institut f{\"u}r Theoretische und Angewandte Physik, Universit{\"a}t 
Stuttgart,  Pfaffenwaldring 57, 70569 Stuttgart, Germany}
\author{Markus Rauscher}
\affiliation{Max-Planck-Institut f{\"u}r
Metallforschung, Heisenbergstr.\ 3, 70569 Stuttgart, Germany, and \\
Institut f{\"u}r Theoretische und Angewandte Physik, Universit{\"a}t 
Stuttgart,  Pfaffenwaldring 57, 70569 Stuttgart, Germany}

\begin{abstract}
We investigate the non-equilibrium fluid structure mediated forces between two colloids driven through a suspension of mutually non-interacting Brownian particles as well as between a colloid and a wall in stationary situations. We solve the Smoluchowski equation in bispherical coordinates as well as with a method of reflections, both in linear approximation for small velocities and numerically for intermediate velocities, and we compare the results to a superposition approximation considered previously. In particular we find an enhancement of the friction (compared to the friction on an isolated particle) for two colloids driven side by side as well as for a colloid traveling along a wall. The friction on tailgating colloids is reduced. Colloids traveling side by side experience a solute induced repulsion while tailgating colloids are attracted to each other.
\end{abstract}


\maketitle

\section{Introduction}\label{sec:intro}

Because colloidal suspensions are ubiquitous in biological as well
as in technological systems, their non-trivial rheological
behavior has been subject to research for a whole century
\cite{einstein06,einstein11}\/. 
Tightly connected to these rheological properties is
the internal dynamics of
these suspensions, 
e.g., sedimentation \cite{klein94}  and diffusion
\cite{ullmann85,phillies85,phillies87,klein96}\/. In the context of microfluidics, the interest in wall
effects in confined geometries has increased. A lot of theoretical effort has been focused on
 hydrodynamic interactions between solute
particles  and walls as well as among solute particles \cite{happel65,kimkarrila91, dhont97}. Also the collective dynamics of hydrodynamically interacting particles has been studied \cite{harris76,felderhof78}. But only recently, the
direct particle-particle interactions and the
resulting fluid structure mediated forces in non-equilibrium have become subject of
theoretical research \cite{dzubiella03b}\/. 

An intuitive picture for this so-called depletion or solvation
force in a non-additive suspension of hard spherical Brownian particles 
has been presented in \cite{asakura54}\/.
Assume two colloidal particles
immersed in a suspension of mutually noninteracting spherical
particles, but with a hard core repulsion with the colloids as
well as with the container walls.
This leads to a forbidden zone around the colloids
which cannot be entered by the Brownian particles. Once the
forbidden zones of the two colloids (or of a colloid and a
container wall)
overlap, the osmotic pressure on the colloids becomes non-uniform,
therefore resulting in a net force. This force has been measured 
in various systems, see e.g., \cite{Bechinger99,helden03,Crocker99}\/.

In a driven system, the
re-distribution of the spherical Brownian particles 
modifies the osmotic pressure leading to long ranged interactions
(long as compared to depletion forces in equilibrium). Some aspects of this
have been studied theoretically for ideal Brownian particles in
\cite{dzubiella03b}\/. The density of the Brownian particles near
the colloids was calculated in a superposition approximation
based on the density in the vicinity of a single colloid
calculated in dynamic density functional theory
\cite{marconi99,marconi00,archer04b}\/. 

In this paper, we calculate this force for a simple
system of mutually non-interacting spherical Brownian particles in
which the DDFT reduces to a simple advected drift-diffusion
equation, i.e., the  Smoluchowski equation. We restrict our analysis of the interaction between two colloids to two paradigmatic cases, namely  
colloids driven side by side and behind each other. In addition, we consider the case of a colloid moving parallel to a solid wall (not discussed in \cite{dzubiella03b}).
In contrast to \cite{dzubiella03b}, we solve the stationary Smoluchowski
equation in bispherical coordinates, analytically for small velocities and numerically for
intermediate velocities. We additionally use a method of reflections as a second alternative to the superposition approximation and discuss the range 
of validity of the different approximations.

\section{Transport equations}
\begin{figure}
\includegraphics[width=1\linewidth]{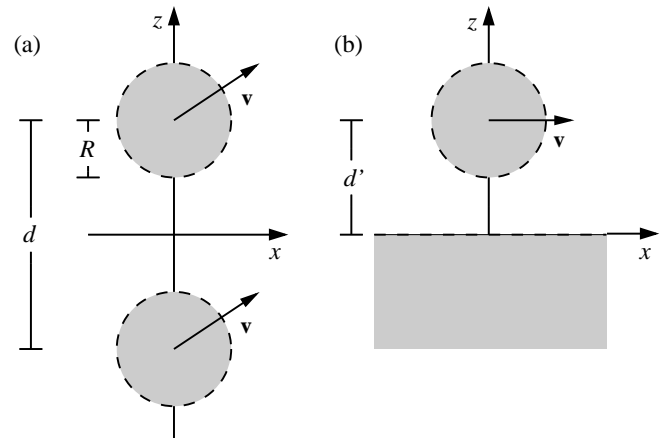}
\caption{\label{fig:0}The two-colloids system (a) and the
colloid-wall system (b). The colloids are driven with velocity
$\vct{v}$. The gray areas $\Gamma_1$ and $\Gamma_2$ are forbidden for the center of the
solute particles, the dashed lines represent their surfaces
$\partial\Gamma_{1/2}$.}
\end{figure}

We consider a stationary driven three component system consisting
of a solvent, Brownian solute particles, e.g., polymer coils, and
(a) two driven colloidal particles at distance $d$, see
Fig.~\ref{fig:0}(a), or (b) one driven colloidal particle at a
distance $d'$ from the surface of the forbidden zone at a planar wall, see Fig.~\ref{fig:0}(b). The
solvent molecules are small compared to any other particle and the
solvent is therefore approximated as homogeneous medium. 
We do not take into account the perturbation of the solvent flow
field by the colloidal particles, i.e., hydrodynamic interactions,  (as studied in \cite{rauscher06b}) for 
three reasons. First, we want to focus on the simplest possible
scenario, second there is no analytical solution available for the
flow field near two spherical particles, and third, we want to
be able to compare our results directly to \cite{dzubiella03b}\/.
We therefore assume a homogeneous solvent flow field in
the frame comoving with the colloids,
$\vct{u}=(-v_x,0,-v_z)$ in the case (a) of two
colloids and $\vct{u}=(-v_x,0,0)$ for case (b), i.e., one colloid
near a wall, see Figs.~\ref{fig:0}(a) and (b), respectively.
The stationary density $\rho(\vct{r})$ of the Brownian particles is then given by
the Smoluchowski equation
\begin{eqnarray}
\frac{\partial\rho}{\partial t}=0=\grad\cdot\vct{j}=-D\,\Delta\rho+\vct{u}\cdot\grad\rho\,,\label{eq:smo}
\end{eqnarray}
where $D$ is the collective diffusion coefficient of the Brownian particles and $\vct{j}$ denotes their probability current density. The hard interaction with the colloids or the
wall leads to no-flux boundary conditions for Eq.~(\ref{eq:smo})
at the surface of the forbidden zone $\partial\Gamma$\/,
see Fig \ref{fig:0}\/. The boundary condition for
Eq.~(\ref{eq:smo}) on $\partial\Gamma$ is 
\begin{eqnarray}
\left.\left(\uvct{n}\cdot\vct{j}\right)\right|_{\partial\Gamma}=0
\label{eq:bc}\,,
\end{eqnarray}
with the surface normal vector $\uvct{n}$ pointing out of the
forbidden zone. 
For colloids with non-overlapping forbidden zones, $\partial\Gamma$ is a
spherical surface of radius $R$, where $R$ is the sum of the
radii of the colloid and the Brownian particles. Far from the
colloids the density should approach the unperturbed density
$\rho_0$\/.

As described in the following, we will solve Eq.~(\ref{eq:smo}) in
bispherical coordinates, which are well adapted to the geometry
under consideration, as long as the forbidden zones do not overlap.
We use a linear expansion for small velocities and we solve
Eq.~(\ref{eq:smo}) numerically for larger velocities. As an
alternative we use the method of reflections, and compare the
results to  the superposition approximation implied in 
\cite{dzubiella03b}\/, where the density near the two colloids is
approximated by the product of the single colloid densities $\rho^{(1)}$ of two colloids with centers at $\vct{R}_1$ and $\vct{R}_2$, 
\begin{eqnarray}
\rho^{(2)}(\vct{r})\approx\rho^{(1)}(\vct{r}-\vct{R}_1)\,\rho^{(1)}(\vct{r}-\vct{R}_2)/\rho_0\,.
\end{eqnarray}

The force on the colloid at $\vct{R}_1$ is given by the integral of the osmotic pressure over the surface of its forbidden zone
$\partial\Gamma_1$. Since there is no mutual interaction between
the Brownian particles, we can use the equation of state of an
ideal gas to calculate the pressure from the local density and get
\begin{eqnarray}
\vct{F}=\int\limits_{\partial\Gamma_1}\,d\vct{S}\,k_B\,T\,\rho\,,
\label{eq:force}
\end{eqnarray}
with surface element $d\vct{S}$.

The solution of Eq.~(\ref{eq:smo}) only depends on the dimensionless velocity
$u^\ast=\frac{|\vct{u}|\,R}{D}$, the Peclet number, and the
dimensionless distances
$d^*=d/R$ and $d'^*=d'/R$\/. Since Eqs.~(\ref{eq:smo}) and (\ref{eq:bc}) are
linear in $\rho$, $\rho$ is proportional to $\rho_0$\/. The
dimensionless force calculated from Eq.~(\ref{eq:force}) 
is given by $F^*=F/(k_B\,T\,\rho_0\,R^2)$\/.

\subsection{Bispherical coordinates}

Bispherical coordinates $(\mu,\eta,\phi)$ are defined via 
\cite{morsefeshbach} 
\begin{eqnarray}
x=\frac{a\,\sin\eta\cos\phi}{\cosh\mu-\cos\eta}\nonumber\,,\\
y=\frac{a\,\sin\eta\sin\phi}{\cosh\mu-\cos\eta}\nonumber\,,\\
z=\frac{a\,\sinh\mu}{\cosh\mu-\cos\eta}\,,
\end{eqnarray}
with  $\mu\in\,\,]\hspace{-0.1cm}-\hspace{-0.1cm}\infty,\infty[\,$,
$\eta\in[0,\pi]$ and $\phi\in[0,2\pi]$\/. The surface $\mu=\pm\mu_0$ is a sphere
of radius $a\,/|\sinh\mu_0|$ centered at $(0,0,\pm\,a\,\coth\mu_0)$,  $\mu=0$
defines the $xy$-plane. For the case (a) of two colloids (see
Fig.~\ref{fig:0}(a)), $\partial\Gamma_{1/2}$ is
given by  $\mu=\pm\mu_0$ and
the distance between the centers in units of $R$ is $d^*=2\,\cosh\mu_0$\/. In
case (b) we choose the surface of the forbidden zone $\Gamma_2$
of the wall (see Fig.~\ref{fig:0}(b)) to lie in the
$xy$-plane and the distance between the colloid and this surface is
$d'^*=\cosh\mu_0$\/. The radius of the spheres  $R$ fixes the scaling factor $a=R\,\sinh\mu_0$\/.

In order to separate variables in the Laplacian in Eq.~(\ref{eq:smo}) we
introduce $h(\vct{r})$ by setting
\begin{equation}
\rho(\vct{r})=\sqrt{\cosh\mu-\cos\eta}\,h(\vct{r})+\rho_0\,.\label{eq:rho}
\end{equation}
Because $\sqrt{\cosh\mu-\cos\eta}\to 0$ far from the colloids
(i.e., $\mu\to 0, \eta\to
0$), the boundary condition $\rho\to\rho_0$ is fulfilled as long as  
$h(\vct{r})$ does not diverge for $|\vct{r}|\to \infty$\/. 
From Eq.~(\ref{eq:smo}) we get for $h(\vct{r})$
\begin{eqnarray}
&&D\,k^2\,\left[\frac{\partial^2h}{\partial\mu^2}+\frac{1}{\sin\eta}\frac{\partial}{\partial\eta}\left(\sin\eta\,\frac{\partial h}{\partial\eta}\right)+\frac{1}{\sin\eta^2}\frac{\partial^2h}{\partial\phi^2}-\frac{1}{4}h\right]\nonumber\\
&&-a\,u_x\,\cos\phi\,\left[-\sinh\mu\,\sin\eta\,\frac{\partial h}{\partial\mu}+(\cosh\mu\,\cos\eta-1)\,\frac{\partial h}{\partial\eta}\nonumber\right.\\
&&\left.-\frac{1}{2}\,\cosh\mu\,\sin\eta\,h-\tan\phi\,\frac{k}{\sin\eta}\,\frac{\partial h}{\partial\phi}\right]\nonumber\\
&&-a\,u_z\,\left[(1-\cosh\mu\,\cos\eta)\,\frac{\partial h}{\partial
\mu}-\frac{1}{2}\,\sinh\mu\,\cos\eta\,h\nonumber\right.\\&&\left.-\sinh\mu\,\sin\eta\,\frac{\partial
h}{\partial \eta}\right]=0\,,\label{langediff}
\end{eqnarray}
with $k=\cosh\mu-\cos\eta$\/. The first term in square brackets, the Laplacian
from Eq.~(\ref{eq:smo}), is diagonalized by the spherical harmonics
$Y_l^m(\eta,\phi)$ \cite{morsefeshbach}. With $\uvct{n}=-\uvct{e}_\mu$ on $\partial\Gamma$ for
$\mu>0$, the boundary condition (\ref{eq:bc}) becomes
\begin{eqnarray}
&&\left\{D\,\left[\frac{1}{2}\,k\,\sinh\mu\,h+k^2\,\frac{\partial h}{\partial \mu}\right]\nonumber\right.\\
&&-a\,u_x\,\left[\left(-\sinh\mu\,\sin\eta\,\cos\phi\right)\,\left(\,h+\frac{\rho_0}{\sqrt{k}}\right)\right]\label{randbed}\\
&&\left.\left.-a\,u_z\,\left[(1-\cosh\mu\,\cos\eta)\left(h+\frac{\rho_0}{\sqrt{k}}\right)\right]\right\}\right|_{\mu=\pm\mu_0}=0\,.\nonumber
\end{eqnarray}
For the two colloids in Fig.~\ref{fig:0}(a), we will only discuss 
the two cases of $\vct{u}$ being
perpendicular and parallel to the axis connecting the two colloids,
while for the case of the colloid in front of the wall in
Fig.~\ref{fig:0}(b), $\vct{u}$ has to be parallel to the wall in
order to allow for stationary solutions.

\subsubsection{Numerical solution}
\begin{figure}
\includegraphics[width=0.49\linewidth]{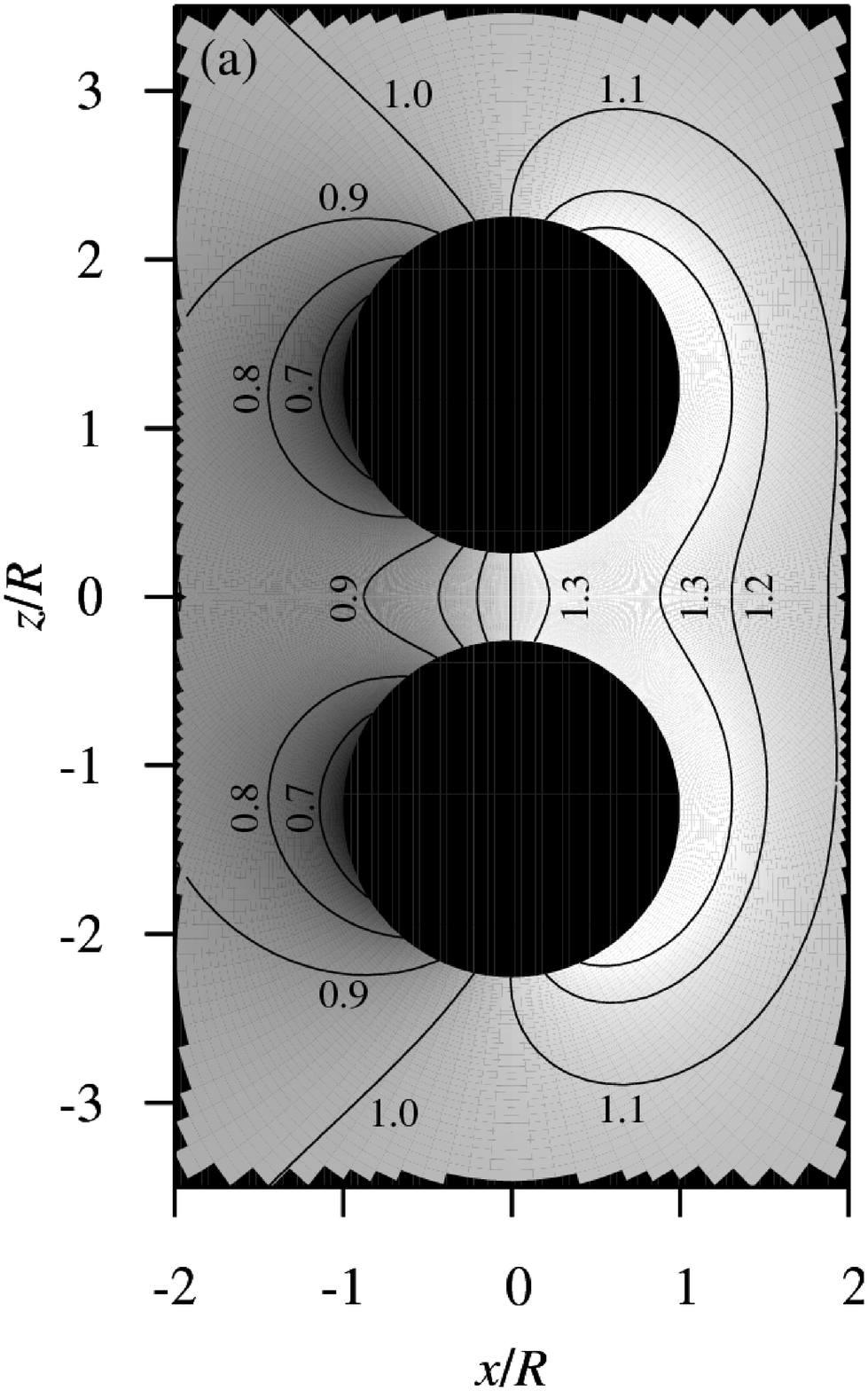}
\includegraphics[width=0.49\linewidth]{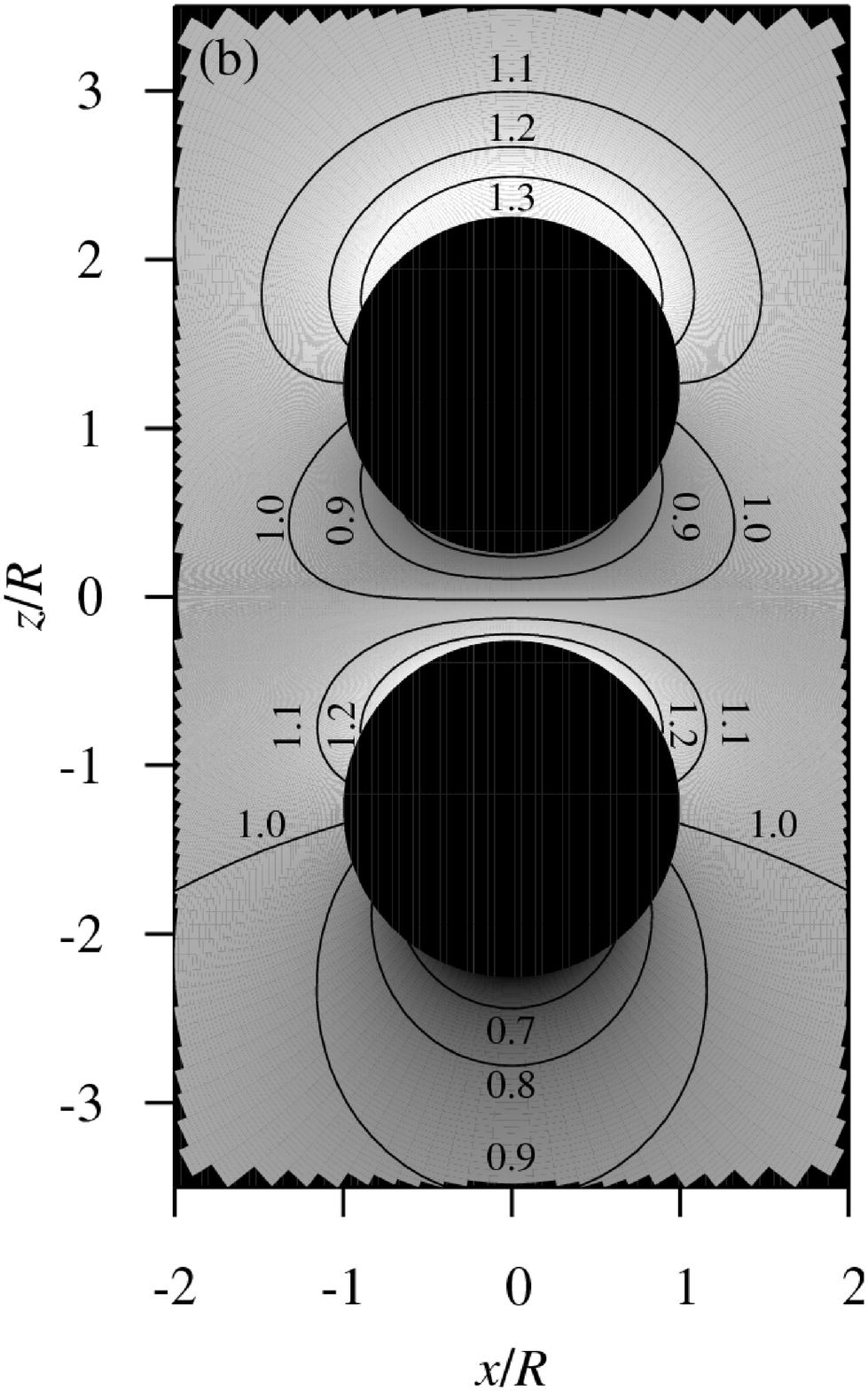}
\caption{\label{fig:2}Contour plots of the stationary density
$\rho/\rho_0$ of solute particles near two driven colloids
for $u^\ast=1$ and $d^*=2.51$. Bright areas correspond to high
densities, dark areas to low densities.
In (a) the colloids are driven
from left to right, the density is enhanced between the colloids
which leads to a repelling force. In (b) they are driven from
bottom to top, the bow wave effect in front of the rear colloid and
the depletion behind the  one in front are reduced, so the friction
forces on the two colloids are reduced as well.}
\end{figure}

The spherical harmonics $Y_l^m(\eta,\phi)$ are eigenfunctions of the angular
part of the Laplace equation for $h(\vct{r})$\/. We therefore expand $h$ in
these functions and project Eq.~(\ref{langediff}) onto them. Taking into account
the symmetry about $\phi=0$, we get $\frac{1}{2}\,N^2+\frac{3}{2}\,N+1$
linear ordinary differential equations for the expansion coefficients
$A_l^m(\mu)$, if the expansion is truncated at $l=N$\/. The same projection
procedure yields the boundary conditions for the $A_l^m(\mu)$ at
$\mu=\pm\mu_0$\/. This set of equations is solved numerically using AUTO~2000
(see http://indy.cs.concordia.ca/auto/), 
a software which solves boundary value problems for systems of
explicit ordinary differential equations by homogeneous
continuation starting from a known solution for a special set of
parameters. Here we started from $u^*=0$, for which
$\rho(\vct{r})=\rho_0$, using $u^*$ as a continuation parameter.
Since $k^2$ multiplies the first term in Eq.~(\ref{langediff}), 
the second derivative of each $A_l^m(\mu)$ appears in more
than one equation. This set of equations has therefore to be solved analytically for these
second derivatives, which restricts us to a maximum order of $N=10$\/. The case
of two colloids driven behind each other has in addition azimuthal symmetry
allowing us to go to $N=15$\/. Dividing Eq.~(\ref{langediff}) by $k^2$ 
before projecting onto spherical harmonics, which yields an
explicit system of equations directly,
would lead to singularities at $\mu,\eta\to0$ and
to additional numerical problems. 
The expansion in spherical harmonics converges badly for $d^*\to2$
(i.e., when the borders of the two forbidden zones get close) and
for large $u^*$\/. In the first case the interval of $\mu$ for
points outside the spheres, where the differential equation is
solved, i.e., $-\mu_0<\mu<\mu_0$, goes to zero. For large $u^*$ sharp variations of $\rho(\vct{r})$
develop near the surfaces of the colloids. For this reason we use
this method only for $u^*\le 2$ and $d^*\ge 2.5$\/.  As an example,
Fig.~\ref{fig:2} shows contour plots of the numerical solutions for
$u^\ast=1$\/. 

\subsubsection{Expansion in $u^*$}

We also seek a solution for small velocities and hence write $h(\vct{r})$ as
a Taylor series expansion in $u^\ast$,
\begin{eqnarray}
h(\vct{r})=h_0(\vct{r})+u^\ast\,h_1(\vct{r})+\mathcal{O}({u^\ast}^2)\,.\label{eq:taylor}
\end{eqnarray}
For $u^\ast\to0$ we require $\rho(\vct{r})\to\rho_0$ and hence $h_0=0$\/. To first order in $u^\ast$, 
Eq.~(\ref{langediff}) reduces to the first term, the Laplace equation for $h_1$, 
\begin{equation}
\frac{\partial^2h_1}{\partial\mu^2}+\frac{1}{\sin\eta}\frac{\partial}{\partial\eta}\left(\sin\eta\,\frac{\partial h_1}{\partial\eta}\right)+\frac{1}{\sin\eta^2}\frac{\partial^2 h_1}{\partial\phi^2}-\frac{1}{4}h_1=0.\label{eq:laph_1}
\end{equation}
The boundary conditions on $\partial \Gamma$ depend on the direction of the flow
field. For colloids traveling side by side we have $u_z^*=0$ leading to 
\begin{multline}
\left.\left(\frac{1}{2}\,k\,\sinh\mu\,h_1+k^2\frac{\partial h_1}{\partial\mu}\right)\right|_{\mu=\pm\mu_0}
\\=\left.\left(a\,\rho_0\,\frac{\sinh\mu\,\sin\eta\,\cos\phi}{\sqrt{k}}\right)\right|_{\mu=\pm\mu_0}\,,
\label{eq:bc1}
\end{multline}
while for the case that one colloid follows the other we have $u_x^*=0$
and we get
\begin{multline}
\left.\left(\frac{1}{2}\,k\,\sinh\mu\,h_1+k^2\,\frac{\partial h_1}{\partial
\mu}\right)\right|_{\mu=\pm\mu_0}\\=-
\left.\left(a\,\rho_0\,\frac{1-\cosh\mu\,\cos\eta}{\sqrt{k}}\right)\right|_{\mu=\pm\mu_0}\label{symm}.
\end{multline}
For the first case, the solution is symmetric with respect to $\mu=0$ and in the
second case antisymmetric.
In both cases the solution is symmetric with respect to $\phi=0$\/. 

The solution of these equations can be expanded in spherical harmonics, see
\cite{morsefeshbach}, leading to an infinite system of linear algebraic equations for the
expansion coefficients, which is solved numerically after truncation at order
$N$\/.
As discussed in the previous subsection, this expansion converges
more slowly for smaller $\mu_0$, i.e., for $d^*\to 2$, so we use
$N=200$ and keep $d^*>2.01$\/. 

\subsection{Method of reflections}\label{sec:refl}

This method has been used successfully to approximate hydrodynamic interactions
 \cite{aguirre73,dhont97}\/. Here we use it to approximate
the density field $\rho(\vct{r})$ for the two colloids with centers at
$\vct{R}_1$ and $\vct{R}_2$\/. Starting point is the solution $\rho^{(1)}$ for a single
colloid, which we calculate by expanding $\rho^{(1)}$ in
spherical harmonics. After truncating the expansion at order $N$ we solve the
resulting system of coupled linear ordinary differential equations numerically
using AUTO~2000\/. For small $u^*$ we get in linear order the simple analytical
result
\begin{equation}
\rho^{(1)}(r,\theta)=\rho_0+u^\ast\,\frac{\rho_0\,R^2}{2\,r^2}\,\cos\theta+\mathcal{O}({u^\ast}^2)
\label{eq:single}\,.
\end{equation}
We then immerse the second colloid  into the density field
$\rho^{(1)}(\vct{r}-\vct{R}_1)$ arising from the first one. The first order
density $\rho_1(\vct{r}-\vct{R}_2)$ reads then
\begin{equation}
\rho_1(\vct{r}-\vct{R}_2)=\rho^{(1)}(\vct{r}-\vct{R}_1)+{\rho_a}_1(\vct{r}-\vct{R}_2)\,,
\end{equation}
where ${\rho_a}_1(\vct{r}-\vct{R}_2)$ solves Eq.~(\ref{eq:smo}) and is chosen
such that $\rho_1(\vct{r}-\vct{R}_2)$ satisfies the boundary condition
(\ref{eq:bc}) at $|\vct{r}-\vct{R}_2|=R$, i.e., on the surface of
the forbidden zone of colloid two\/.
Re-immersing the first colloid at $\vct{R}_1$ into the density field
$\rho_1(\vct{r}-\vct{R}_2)$ yields the second order density 
\begin{equation}
\rho_2(\vct{r}-\vct{R}_1)=\rho_1(\vct{r}-\vct{R}_2)+{\rho_a}_2(\vct{r}-\vct{R}_1)\,,
\end{equation}
where ${\rho_a}_2(\vct{r}-\vct{R}_1)$ adjusts the boundary condition at
$|\vct{r}-\vct{R}_1|=R$\/. This procedure can be repeated and the
iteration is assumed to converge to the exact solution for two
colloids.
The approximated solutions $\rho_1, \rho_2, \dots$ always solve the Smoluchowski
equation (\ref{eq:smo}) and additionally satisfy the boundary condition on one
of the colloids. 

In first order of $u^*$ the force on the colloids resulting from
$\rho_1$ can be calculated analytically. For higher orders in
reflections as well as for large $u^*$, the changes of coordinates
from one colloid to the other in the reflection procedure has to be
performed numerically.

\section{Colloids driven side by side}
In this section we study the case of two colloids driven side by
side through a solution of solute particles. We start with an
expansion in the dimensionless flow velocity $u^\ast$, where all
the introduced approaches can be evaluated analytically,
and later discuss numerical solutions for higher velocities.

\subsection{Expansion in $u^\ast$}
\subsubsection{Bispherical coordinates}
\label{secIIa1}
Due to the symmetries of the system and because the right hand side of
Eq.~(\ref{eq:bc1}) is proportional to $\cos\phi$ 
the expansion of $h_1$ simplifies to
\begin{equation}
h_1=\sum_{l=0}^\infty\,C_l^1\,\cosh\left[\left(l+\frac{1}{2}\right)\mu\right]\,P_l^1(\cos\eta)\,\cos \phi\label{eq:h_12}\,.
\end{equation}
Projecting Eq.~(\ref{eq:bc1}) on $P_l^1(\cos\eta)\,\cos\phi$, one
gets a linear system of equations for the $C_l^1$\/. Here we need
the following expansion (see \cite{morsefeshbach})
 \begin{equation}
 \frac{1}{\sqrt{\cosh\mu-\cos\eta}}=\sqrt{2}\,\sum\limits_{l=0}^{\infty}e^{-(l+\frac{1}{2})|\mu|}\,P_l(\cos\eta)\,,
 \label{eq:ap:1}
 \end{equation}
in order to obtain with a recursion relation for the Legendre
polynomials 
 \begin{align}
 \frac{\sin\eta\cos\phi}{\sqrt{\cosh\mu-\cos\eta}}=&\sqrt{2}\,\sum\limits_{l=0}^{\infty}e^{-(l+\frac{1}{2})|\mu|}\,\frac{1}{2l+1}\\
 &\times\bigl[P_{l-1}^1(\cos\eta)-P_{l+1}^1(\cos\eta)\bigr]\,\cos\phi\nonumber\,,
\end{align}
which is used for the projection of the right side of
Eq.~(\ref{eq:bc1}) onto $P_l^1(\cos\eta)\,\cos\phi$\/.

To first order in $u^\ast$, the force defined in Eq.~(\ref{eq:force}) in
$z$-direction vanishes by symmetry, so there is no solute particle
mediated interaction between the two colloids. A term linear in
$u^\ast$ would change sign if $u^\ast$ changes sign. With the same
type of argument one can show that there is no hydrodynamic
interaction between the colloids in this setup. The friction force
parallel to the flow $\vct{u}$, i.e., in $x$-direction, is
enhanced compared to a single colloid, since more solute
particles are collected in front of two colloids. This force as a
function of the distance $d$ of the centers of the colloids is
depicted in Fig.~\ref{fig:1a}\/. 

\subsubsection{Method of reflections}
\begin{figure}
\includegraphics[angle=270,width=1\linewidth]{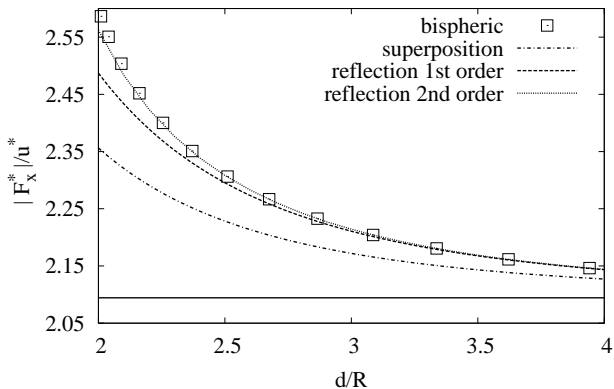}
\caption{\label{fig:1a}Friction force $F^\ast$ of ideal solute particles on one of the two colloids driven side by side to first order in $u^\ast$. The horizontal line at $|F_x^\ast|/u^\ast=\frac{2\,\pi}{3}\approx2.1$ is the limit for $d\to\infty$. The same force
applies to a colloid which is driven parallel to a planar wall at distance $d'=d/2$.}
\end{figure}

In first order in $u^\ast$, Eq.~(\ref{eq:smo}) reduces to the
Laplace equation. The solutions which do not diverge for
$r\to\infty$ can be expanded in terms of spherical harmonics as
follows
\begin{equation}
\rho(r,\theta,\phi)=\sum_{l,m}D_l^m\,r^{-(l+1)}\,Y_l^m(\theta,\phi)\,.
\end{equation}
The adjusted densities 
can be expanded in the same way.
Starting with Eq.~(\ref{eq:single}) for a single colloid, we get
for the adjusted density on the second one
\begin{equation}
\rho_{a1}(\vct{r}-\vct{R}_2)=u^\ast\rho_0\left(\frac{1}{2}+\frac{R^3}{4\,d^3}\right)\,\frac{R^2}{(\vct{r}-\vct{R}_2)^2}\,\cos\theta+\dots\,,
\end{equation}
where $\theta$ is the angle between $\vct{v}$ and
$\vct{r}-\vct{R}_2$\/.  The dots represent higher order terms in $l$,
which are important for the second reflection but do not
contribute to the friction force in first order. The density
$\rho_1$ on the surface of the forbidden zone of the second colloid is thus given by
\begin{equation}
\rho_1(\vct{r}-\vct{R}_2)|_{|\vct{r}-\vct{R}_2|=R}=\rho_0+u^\ast\rho_0\left(\frac{1}{2}+\frac{3\,R^3}{4\,d^3}\right)\,\cos\theta+\dots\,.
\end{equation}
The force on the colloid in first order of reflection is hence given by 
\begin{equation}
\label{eq:refl1}
F_x^{r}=-u^\ast\,R^2\,\rho_0\,k_BT\,\frac{4\,\pi}{3}\,\left(\frac{1}{2}+\frac{3\,R^3}{4\,d^3}\right)+\mathcal{O}({u^\ast}^2)\,,
\end{equation}
which is also shown in Fig.~\ref{fig:1a}. The second reflection
changes the force to order $\mathcal{O}\left(\frac{1}{d^4}\right)$
and higher, but not the term $\propto \frac{1}{d^3}$\/. Hence the
first order reflection gives already the exact asymptotic form for
$d\to\infty$\/. This can be seen in Fig.~\ref{fig:1a}, the graphs
for first and second reflection have the same asymptotic behavior.
The second reflection shows good agreement with the linear
bispherical results.
\subsection{Discussion and numerical solution}

Fig.~\ref{fig:1a} compares the drag force for small velocities
obtained in bispherical coordinates and with the method of
reflections with the result from the superposition approximation. 
With the solution for a single colloid in
first order in $u^\ast$ given by Eq.~(\ref{eq:single})
we get in the superposition approximation the drag force 
\begin{equation}
F_x^{s}=-u^\ast\,R^2\,\rho_0\,k_BT\,\frac{4\,\pi}{3}\,\left(\frac{1}{2}+\frac{R^3}{2\,d^3}\right)+\mathcal{O}({u^\ast}^2)\,.
\end{equation}
By construction,
both the expression obtained with the method of reflection in
Eq.~(\ref{eq:refl1}) as well as the expression above converge for
$d\to\infty$ to the force on a single colloid. % by construction. 
Both have the same power law dependence $\propto d^{-3}$, but with
a different prefactor. For large $d$ already the first order
reflections approximate the linearized bispherical result from
Sec.~\ref{secIIa1} well.
For smaller $d$ the second order reflection result is needed.

\begin{figure}
\includegraphics[angle=270,width=1\linewidth]{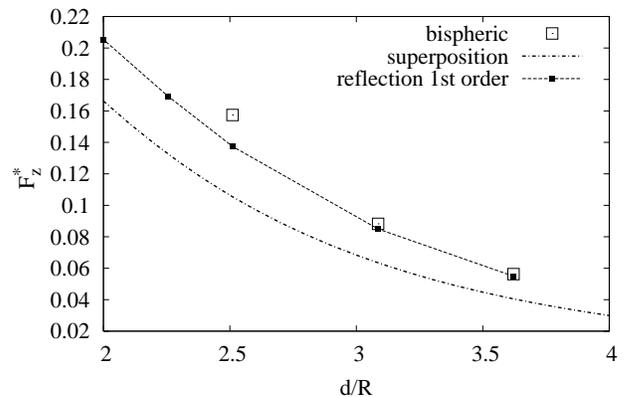}
\caption{\label{fig:1b}Repelling force $F^\ast$ of ideal solute
particles between two colloids driven side by side with velocity
$u^\ast=1$. The same force acts on a colloid which is driven
parallel to a planar wall at distance $d'=d/2$.}
\end{figure}

For higher velocities, the force between the two colloids is non
vanishing and repulsive. Note that this is the only repulsive force
(apart from possible direct interactions between the colloids)
since there are no hydrodynamic interactions. 
It increases with decreasing distance
$d$, see Fig.~\ref{fig:1b}\/. The force calculated with the method
of reflections to first order and the numerical result obtained in
bispherical coordinates agree well for large $d$, suggesting that
the first order reflections give again the correct asymptotic
behavior for $d\to\infty$\/.

\begin{figure}
\includegraphics[angle=270,width=1\linewidth]{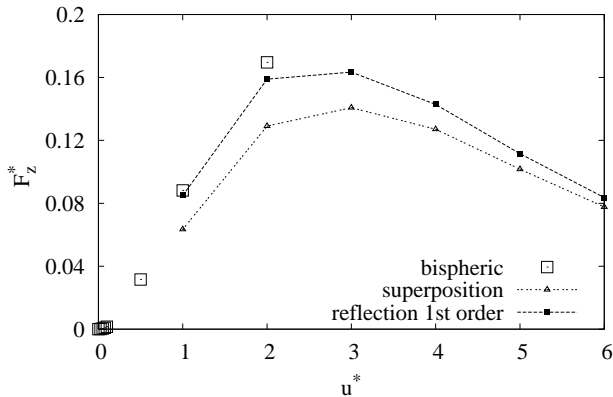}
\caption{\label{fig:1c}Repelling force $F^\ast$ of ideal solute
particles between two colloids driven side by side at distance
$d/R=3.1$. The same force acts on a colloid which is driven
parallel to a planar wall at distance $d'/R=1.54$.}
\end{figure}

This repelling force for fixed distance shows an interesting
dependence on the velocity. For small velocities, it increases
proportional to ${u^\ast}^2$, has a maximum and goes to zero for
$u^\ast\to\infty$, see Fig.~\ref{fig:1c}\/. The reason is, that
the force is a symmetric function of $u^*$ and that the range of
interaction goes to zero for $u^*\to \infty$\/. For large $u^*$
the thickness of the region in which the density is modified by
the presence of the colloid decreases, such that the other colloid
is immersed in a less perturbed environment. This is also the
reason why first order reflections as well as the superposition
approximation are expected to work well. %give a good result.
Here, for the superposition we use the numerical solution for a
single colloid which is also used as a starting distribution for
the method of reflections.


\section{Colloids driven behind each other}
Here we investigate the case of two colloids driven behind each
other through the solution of Brownian particles. Again, we
evaluate the approaches analytically in first order in $u^\ast$ as
well as numerically for higher velocities.
\subsection{Expansion in $u^\ast$}
\subsubsection{Bispherical coordinates}
In the case considered here, the flow is parallel to the $z$-axis,
hence $u_x=0$ in Eqs. (\ref{langediff}) and (\ref{randbed}). The
system
possesses azimuthal symmetry, so the $Y_l^m(\eta,\phi)$ reduce to
the Legendre polynomials $P_l(\cos\eta)$\/. As in
Eq.~(\ref{eq:taylor}) we write $h(\vct{r})$ as a Taylor series in
$u^\ast$ with $h_0=0$\/. $h_1$ is again the solution of
Eq.~(\ref{eq:laph_1}) and given by 
\begin{eqnarray}
h_1=\sum_{l=0}^\infty\,S_l\,\sinh\left[\left(l+\frac{1}{2}\right)\mu\right]\,P_l(\cos\eta)\,.
\end{eqnarray}
Projecting Eq.~(\ref{symm}) at $\mu=\mu_0$ on the $P_l(\cos\eta)$,
one obtains a linear set of equations for the coefficients $S_l$,
which can  be solved after truncating at $l=N$\/. The term
involving the square root of $k$ is again projected via
Eq.~(\ref{eq:ap:1}) and a recursion relation.

\begin{figure}
\includegraphics[angle=270,width=1\linewidth]{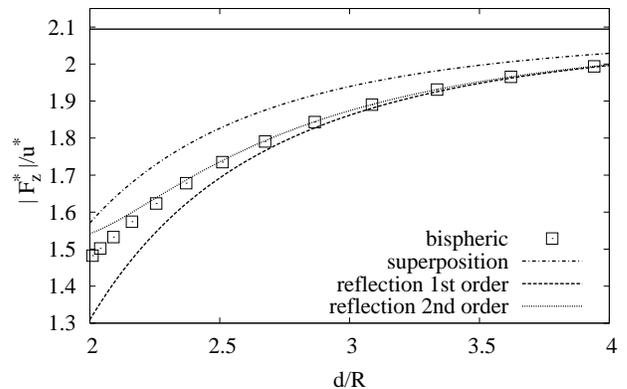}
\caption{\label{fig:3a}Friction force $F^\ast$ induced by the solute
particles on one of two colloids driven behind each other to first
order in $u^\ast$. The horizontal line is the limit for
$d\to\infty$, i.e., the friction force for a single colloid.
}
\end{figure}

The friction force as calculated from Eq.~(\ref{eq:force}) is
reduced compared to the force on a single colloid, see
Fig.~\ref{fig:3a}\/. To first order in $u^\ast$, this friction
force (as well as the hydrodynamic drag) is identical for both colloids 
by symmetry. The reason is that $\rho(x,y,z,u^*) =
\rho(x,y,-z,-u^*)$ and therefore the coefficient of the linear term
in an expansion in $u^*$ has to be antisymmetric in $z$\/.

\subsubsection{Method of reflections}
The density distribution in front and behind a single colloid is
different and therefore there exist two different first order
solutions to the method of reflections, namely for immersing the
second colloid in front or behind the first one.  However, the
forces on the colloids depend only on the term in the density on
the colloid proportional to $\cos\theta$\/. In first order
 of reflection this term is the same for both solutions and it is given by 
\begin{equation}
u^\ast\rho_0\left(\frac{1}{2}-\frac{3\,R^3}{2\,d^3}\right)\,\cos\theta.
\end{equation}
The resulting force is then given by
\begin{equation}
F_z^r=-u^\ast\,R^2\,\rho_0\,k_BT\,\frac{4\,\pi}{3}\,\left(\frac{1}{2}-\frac{3\,R^3}{2\,d^3}\right)+\mathcal{O}({u^\ast}^2)\,,
\end{equation}
which is also shown in Fig.~\ref{fig:3a}\/. 
Additionally one can show that to all orders of reflections, the
force on the colloids does not depend on whether one starts the
reflection method with the first or the second colloid, reflecting
the fact, that in first order in $u^*$, the friction forces on the
two colloids are identical. 
As in the previous case of two colloids moving side by side, the
asymptotic behavior of $F_z^r$ for $d\to\infty$ is not changed by
the second reflection.
\subsection{Discussion and numerical solution}
Fig.~\ref{fig:3a} compares the drag force for small velocities
obtained in bispherical coordinates and with the method of
reflections with the result from the superposition approximation.
For the superposition we use again Eq.~(\ref{eq:single}) keeping
only terms linear in $u^\ast$ in the product of the densities.  The
resulting force is given by
\begin{equation}
F_z^s=-u^\ast\,R^2\,\rho_0\,k_BT\,\frac{4\,\pi}{3}\,\left(\frac{1}{2}-\frac{R^3}{d^3}\right)+\mathcal{O}({u^\ast}^2)\,.
\end{equation}
To first order in $u^*$, we find the same features as in the
side-by-side case.  The drag forces obtained by the method of
reflections and the superposition approximation converge for large
$d$ to the force on a single colloid with the same  power law
$d^{-3}$ but with a different prefactor. The first order reflection
agrees well with the linearized bispherical results for large $d$,
for small $d$ higher order reflections are needed.

\begin{figure}
\includegraphics[angle=270,width=1\linewidth]{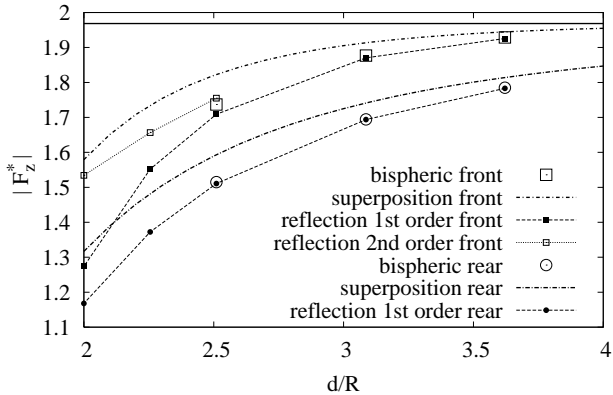}
\caption{\label{fig:3b}Friction forces $F^\ast$ induced by solute
particles on the two colloids driven behind each other for
$u^\ast=1$\/. The limit of $d\to\infty$ (i.e., the force on a
single colloid) is given by $|F_z^\ast|=1.97$.}
\end{figure}

For higher velocities, the friction on the front colloid is larger
than on the rear one which is shielded, see Fig.~\ref{fig:3b}\/.
For large $d$ the forces obtained with first order reflection agree
well with the forces calculated from the numerical solution in
bispherical coordinates. Here, in order to calculate the force on
the rear colloid, we start with the colloid in front and vice
versa. For smaller $d$, the agreement is also good for the forces
on the rear colloid, but for the front colloid higher order
reflections are necessary. The reason is, that the rear colloid
influences the density distribution on the front colloid much less
than vice versa. 

The fact that the friction force on both colloids is reduced as
compared to the case of a single colloid can be understood in the
following way. The bow wave of particles accumulated in front of
the rear colloid push the front colloid, while the rear colloid
moves in a region of reduced particle density and experiences
less friction. 

\section{Colloid near a wall}
\begin{figure}
\includegraphics[width=0.49\linewidth]{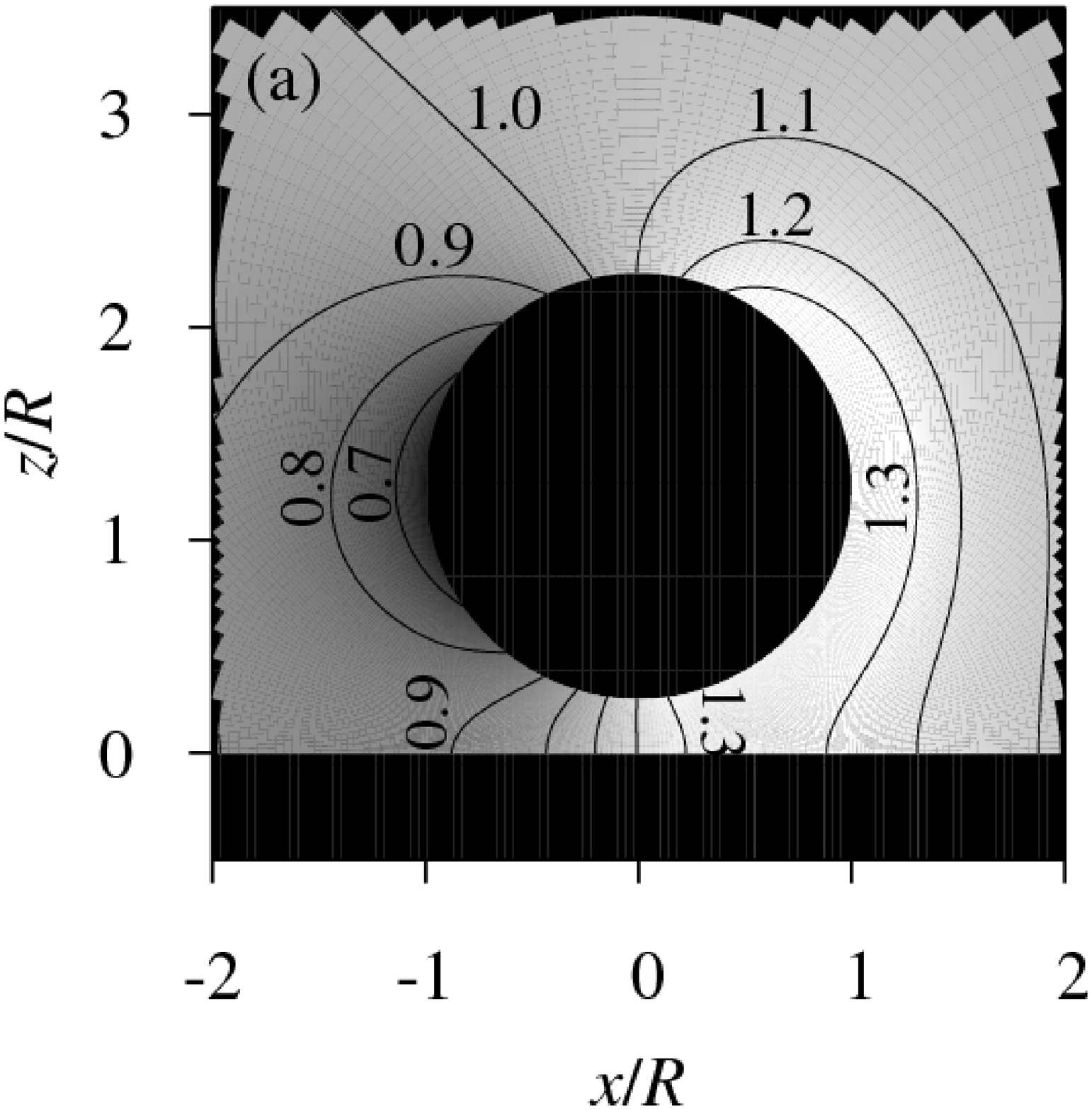}
\includegraphics[width=0.49\linewidth]{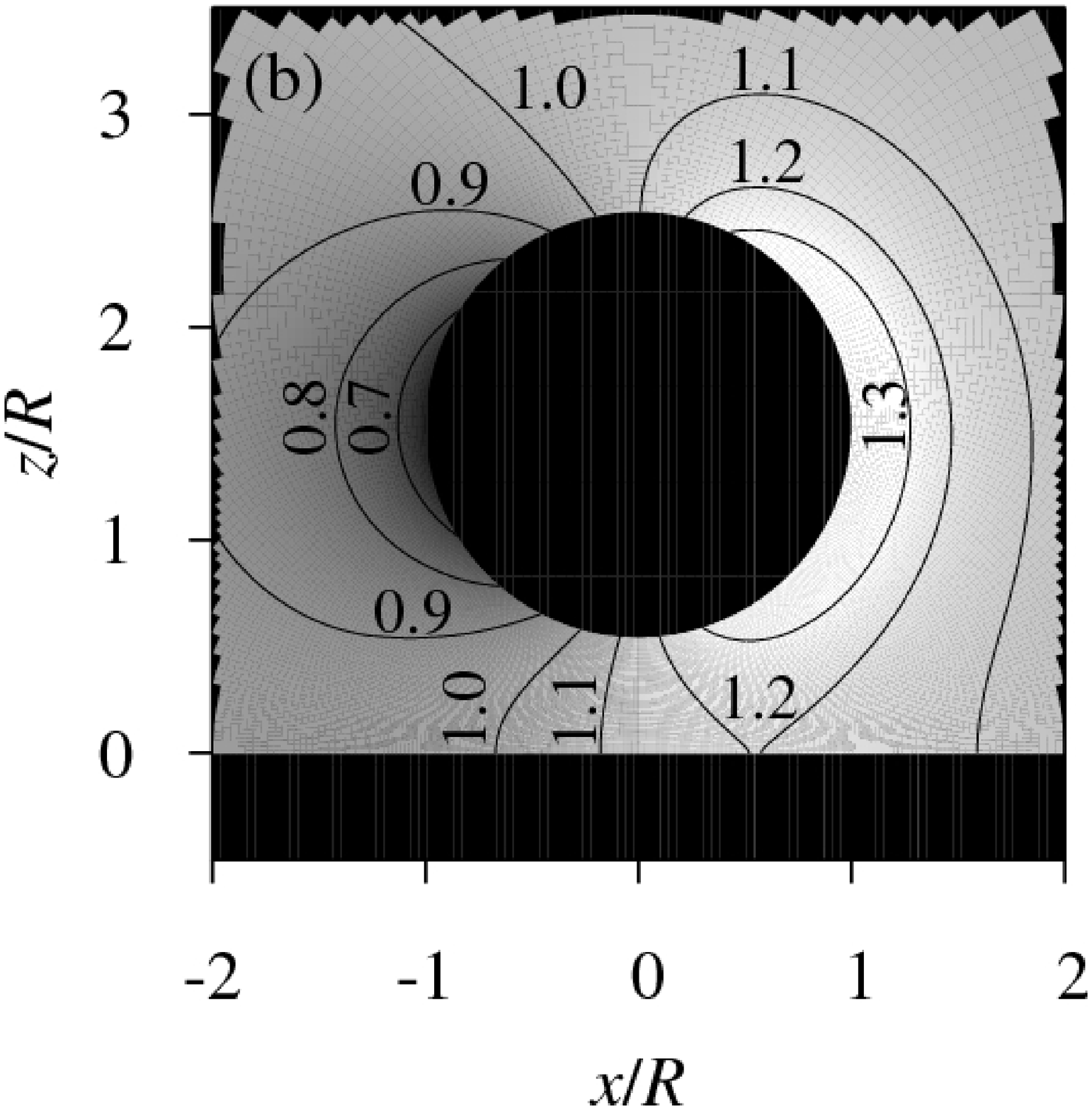}
\caption{\label{fig:4}Contour plots of the stationary density
$\rho/\rho_0$ of ideal solute particles for a colloid driven from
left to right parallel to a wall at distance (a) $d'/R=1.26$ and
(b) $d'/R=1.54$. (a) is exactly the upper half of Fig.
\ref{fig:2}(a). 
}
\end{figure}

Here we consider the case of a colloid, which is driven parallel to a planar
wall. As mentioned above, the surface $\mu=0$ is the $xy$-plane which we choose
to be the surface of the forbidden zone in front of the wall. So the boundary
condition Eq.~(\ref{randbed}) must be evaluated at $\mu=\mu_0$ and $\mu=0$\/.
The flow field $\vct{u}$ is parallel to the wall, $\vct{u}=u_x\,\uvct{e}_x$ and
Eq.~(\ref{randbed}) at $\mu=0$ becomes 
\begin{equation}
\left.\frac{\partial h}{\partial\mu}\right|_{\mu=0}=0\,.
\end{equation}
This condition is fulfilled by the solution for the side-by-side case since it
is symmetric about $\mu=0$\/. The condition at $\mu=\mu_0$ is also satisfied. So
the solutions for these two cases are identical including the forces on the
colloid, i.e., it is repelled by the wall and the friction is enhanced as
compared to the case without wall, see Figs.~\ref{fig:1a} to \ref{fig:1c}\/. 
This is reminiscent of the method of images used in electrostatics.
Fig.~\ref{fig:4} shows contour plots of the stationary density distribution for
two values of $d'/R$\/.
\section{Discussion}\label{sec:dis}

We find repelling solute particle mediated forces between two colloids driven side by side which vanishes to first order in the velocity. This force as a function of the drift velocity shows a maximum and goes to zero at infinity since the range of the interactions goes to zero in this limit. The friction on the colloids is enhanced compared to a single one. Via the method of images the same applies to a colloid which is driven parallel to a planar wall, i.e., it is repelled and its friction is enhanced. When the colloids are driven behind each other, the friction is reduced and to first order in the velocity identical for both colloids. For higher velocities, the first colloid feels a larger friction than the second one. 

For the behind-each-other case, we find reasonable agreement
between the results in bispherical coordinates and the
superposition approximation. For the side-by-side case the
agreement is less good. Our explanation is causality. Due to the
motion of the colloids,  perturbations of the density behind a
colloid hardly influence the density around it, but perturbations
of the particle density in front of the colloid do influence the
density on its surface significantly. When the colloids are driven
behind each other, the first one is almost decoupled from the
second one and back-coupling effects between the colloids, which
are neglected by the superposition approximation, are less
important. So this approximation works well in this case. For the
same reason, the first order reflection on the rear colloid already
agrees very well with the bispherical results for $u^\ast=1$. For
the side-by-side case, back-coupling effects are more important
since the colloids are on the same level and the superposition
approximation gives worse results. The flow limited persistence
time for one of the solute particles between the two colloids is
given by 
$t=\frac{R}{|\vct{u}|}$\/. For large velocities this time goes to
zero and a particle has no time to diffuse from one colloid to the
other. Therefore back-coupling effects become negligible. The consequences of
this can be observed in Fig.~\ref{fig:1c}: the
superposition-results approach the first order reflection in this
limit.

Comparing the forces on the two colloids driven through a
suspension of ideal Brownian particles obtained in bispherical
coordinates, with the method of reflections, and in the
superposition approximation we find a qualitative agreement between
all three methods in the range in which they can be applied.  All
methods are applicable for large distances, however, the
superposition approximation yields the wrong prefactor in the
asymptotics. For large $u^*$ the numerical algorithm using
bispherical coordinates shows bad convergence but the superposition
approximation and the method of reflections show the same
qualitative behavior.  In summary we find that the method of
reflections is more reliable than the superposition approximation.
The numerical solution using bispherical coordinates has only a
restricted range of applicability. In order to further test the
method of reflections a finite element scheme could be used, which
would allow for localized refinement of the discretization right at
the surface of the forbidden zones. 
This would also allow to study the dynamics of the system.

\begin{acknowledgments}
We thank S. Dietrich for support and fruitful discussions.
M.R. acknowledges funding by the Deutsche Forschungsgemeinschaft within the
priority program SPP~1164 ``Micro- and Nanofluidics'' under grant
number RA~1061/2-1.
\end{acknowledgments}

%\bibliography{references}

\begin{thebibliography}{23}
\expandafter\ifx\csname natexlab\endcsname\relax\def\natexlab#1{#1}\fi
\expandafter\ifx\csname bibnamefont\endcsname\relax
  \def\bibnamefont#1{#1}\fi
\expandafter\ifx\csname bibfnamefont\endcsname\relax
  \def\bibfnamefont#1{#1}\fi
\expandafter\ifx\csname citenamefont\endcsname\relax
  \def\citenamefont#1{#1}\fi
\expandafter\ifx\csname url\endcsname\relax
  \def\url#1{\texttt{#1}}\fi
\expandafter\ifx\csname urlprefix\endcsname\relax\def\urlprefix{URL }\fi
\providecommand{\bibinfo}[2]{#2}
\providecommand{\eprint}[2][]{\url{#2}}

\bibitem[{\citenamefont{Einstein}(1906)}]{einstein06}
\bibinfo{author}{\bibfnamefont{A.}~\bibnamefont{Einstein}},
  \bibinfo{journal}{Annalen der Physik} \textbf{\bibinfo{volume}{19}},
  \bibinfo{pages}{289} (\bibinfo{year}{1906}).

\bibitem[{\citenamefont{Einstein}(1911)}]{einstein11}
\bibinfo{author}{\bibfnamefont{A.}~\bibnamefont{Einstein}},
  \bibinfo{journal}{Annalen der Physik} \textbf{\bibinfo{volume}{34}},
  \bibinfo{pages}{1911} (\bibinfo{year}{1911}).

\bibitem[{\citenamefont{Klein and Nagele}(1994)}]{klein94}
\bibinfo{author}{\bibfnamefont{R.}~\bibnamefont{Klein}} \bibnamefont{and}
  \bibinfo{author}{\bibfnamefont{G.}~\bibnamefont{Nagele}},
  \bibinfo{journal}{Nuovo Cimento D} \textbf{\bibinfo{volume}{16}},
  \bibinfo{pages}{963} (\bibinfo{year}{1994}).

\bibitem[{\citenamefont{Ullmann et~al.}(1985)\citenamefont{Ullmann, Ullmann,
  Lindner, and Phillies}}]{ullmann85}
\bibinfo{author}{\bibfnamefont{G.~S.} \bibnamefont{Ullmann}},
  \bibinfo{author}{\bibfnamefont{K.}~\bibnamefont{Ullmann}},
  \bibinfo{author}{\bibfnamefont{R.~M.} \bibnamefont{Lindner}},
  \bibnamefont{and} \bibinfo{author}{\bibfnamefont{G.~D.~J.}
  \bibnamefont{Phillies}}, \bibinfo{journal}{J. Phys. Chem.}
  \textbf{\bibinfo{volume}{89}}, \bibinfo{pages}{692} (\bibinfo{year}{1985}).

\bibitem[{\citenamefont{Phillies et~al.}(1985)\citenamefont{Phillies, Ullmann,
  Ullmann, and Lin}}]{phillies85}
\bibinfo{author}{\bibfnamefont{G.~D.~J.} \bibnamefont{Phillies}},
  \bibinfo{author}{\bibfnamefont{G.~S.} \bibnamefont{Ullmann}},
  \bibinfo{author}{\bibfnamefont{K.}~\bibnamefont{Ullmann}}, \bibnamefont{and}
  \bibinfo{author}{\bibfnamefont{T.-H.} \bibnamefont{Lin}},
  \bibinfo{journal}{J. Chem. Phys.} \textbf{\bibinfo{volume}{82}},
  \bibinfo{pages}{5242} (\bibinfo{year}{1985}).

\bibitem[{\citenamefont{Phillies et~al.}(1987)\citenamefont{Phillies, Malone,
  Ullmann, Ullmann, Rollings, and Yu}}]{phillies87}
\bibinfo{author}{\bibfnamefont{G.~D.~J.} \bibnamefont{Phillies}},
  \bibinfo{author}{\bibfnamefont{C.}~\bibnamefont{Malone}},
  \bibinfo{author}{\bibfnamefont{K.}~\bibnamefont{Ullmann}},
  \bibinfo{author}{\bibfnamefont{G.~S.} \bibnamefont{Ullmann}},
  \bibinfo{author}{\bibfnamefont{J.}~\bibnamefont{Rollings}}, \bibnamefont{and}
  \bibinfo{author}{\bibfnamefont{L.-P.} \bibnamefont{Yu}},
  \bibinfo{journal}{Macromolecules} \textbf{\bibinfo{volume}{20}},
  \bibinfo{pages}{2280} (\bibinfo{year}{1987}).

\bibitem[{\citenamefont{Klein and Nagele}(1996)}]{klein96}
\bibinfo{author}{\bibfnamefont{R.}~\bibnamefont{Klein}} \bibnamefont{and}
  \bibinfo{author}{\bibfnamefont{G.}~\bibnamefont{Nagele}},
  \bibinfo{journal}{Current Opinion in Colloid \& Interface Science}
  \textbf{\bibinfo{volume}{1}}, \bibinfo{pages}{4} (\bibinfo{year}{1996}).

\bibitem[{\citenamefont{Dhont}(1997)}]{dhont97}
\bibinfo{author}{\bibfnamefont{J.~K.~G.} \bibnamefont{Dhont}},
  \emph{\bibinfo{title}{An Introduction to Dynamics of Colloids}},
  vol.~\bibinfo{volume}{II} of \emph{\bibinfo{series}{Studies in Interface
  Science}} (\bibinfo{publisher}{Elsevier}, \bibinfo{address}{Amsterdam},
  \bibinfo{year}{1997}).

\bibitem[{\citenamefont{Happel and Brenner}(1965)}]{happel65}
\bibinfo{author}{\bibfnamefont{J.}~\bibnamefont{Happel}} \bibnamefont{and}
  \bibinfo{author}{\bibfnamefont{H.}~\bibnamefont{Brenner}},
  \emph{\bibinfo{title}{Low {R}eynolds Number Hydrodynamics}}
  (\bibinfo{publisher}{Prentice Hall}, \bibinfo{address}{New Jersey},
  \bibinfo{year}{1965}).

\bibitem[{\citenamefont{Kim and Karrila}(1991)}]{kimkarrila91}
\bibinfo{author}{\bibfnamefont{S.}~\bibnamefont{Kim}} \bibnamefont{and}
  \bibinfo{author}{\bibfnamefont{S.~J.} \bibnamefont{Karrila}},
  \emph{\bibinfo{title}{Microhydrodynamics : principles and selected
  applications}} (\bibinfo{publisher}{Butterworth-Heinemann},
  \bibinfo{address}{Boston}, \bibinfo{year}{1991}).

\bibitem[{\citenamefont{Harris}(1976)}]{harris76}
\bibinfo{author}{\bibfnamefont{S.}~\bibnamefont{Harris}}, \bibinfo{journal}{J.
  Phys. A: Math. Gen.} \textbf{\bibinfo{volume}{9}}, \bibinfo{pages}{1895}
  (\bibinfo{year}{1976}).

\bibitem[{\citenamefont{Felderhof}(1978)}]{felderhof78}
\bibinfo{author}{\bibfnamefont{B.~U.} \bibnamefont{Felderhof}},
  \bibinfo{journal}{J. Phys. A: Math. Gen.} \textbf{\bibinfo{volume}{11}},
  \bibinfo{pages}{929} (\bibinfo{year}{1978}), \bibinfo{note}{times Cited:
  269}.

\bibitem[{\citenamefont{Dzubiella et~al.}(2003)\citenamefont{Dzubiella,
  L{\"o}wen, and Likos}}]{dzubiella03b}
\bibinfo{author}{\bibfnamefont{J.}~\bibnamefont{Dzubiella}},
  \bibinfo{author}{\bibfnamefont{H.}~\bibnamefont{L{\"o}wen}},
  \bibnamefont{and} \bibinfo{author}{\bibfnamefont{C.~N.} \bibnamefont{Likos}},
  \bibinfo{journal}{Phys. Rev. Lett.} \textbf{\bibinfo{volume}{91}},
  \bibinfo{pages}{248301} (\bibinfo{year}{2003}),
  \bibinfo{note}{cond-mat/0306069}.

\bibitem[{\citenamefont{Asakura and Oosawa}(1954)}]{asakura54}
\bibinfo{author}{\bibfnamefont{S.}~\bibnamefont{Asakura}} \bibnamefont{and}
  \bibinfo{author}{\bibfnamefont{F.}~\bibnamefont{Oosawa}},
  \bibinfo{journal}{J. Chem. Phys.} \textbf{\bibinfo{volume}{22}},
  \bibinfo{pages}{1255} (\bibinfo{year}{1954}).

\bibitem[{\citenamefont{Bechinger et~al.}(1999)\citenamefont{Bechinger,
  Rudhardt, Leiderer, Roth, and Dietrich}}]{Bechinger99}
\bibinfo{author}{\bibfnamefont{C.}~\bibnamefont{Bechinger}},
  \bibinfo{author}{\bibfnamefont{D.}~\bibnamefont{Rudhardt}},
  \bibinfo{author}{\bibfnamefont{P.}~\bibnamefont{Leiderer}},
  \bibinfo{author}{\bibfnamefont{R.}~\bibnamefont{Roth}}, \bibnamefont{and}
  \bibinfo{author}{\bibfnamefont{S.}~\bibnamefont{Dietrich}},
  \bibinfo{journal}{Phys. Rev. Lett.} \textbf{\bibinfo{volume}{83}},
  \bibinfo{pages}{3960} (\bibinfo{year}{1999}).

\bibitem[{\citenamefont{Helden et~al.}(2003)\citenamefont{Helden, Roth,
  Koenderink, Leiderer, and Bechinger}}]{helden03}
\bibinfo{author}{\bibfnamefont{L.}~\bibnamefont{Helden}},
  \bibinfo{author}{\bibfnamefont{R.}~\bibnamefont{Roth}},
  \bibinfo{author}{\bibfnamefont{G.~H.} \bibnamefont{Koenderink}},
  \bibinfo{author}{\bibfnamefont{P.}~\bibnamefont{Leiderer}}, \bibnamefont{and}
  \bibinfo{author}{\bibfnamefont{C.}~\bibnamefont{Bechinger}},
  \bibinfo{journal}{Phys. Rev. Lett.} \textbf{\bibinfo{volume}{90}},
  \bibinfo{pages}{048301} (\bibinfo{year}{2003}).

\bibitem[{\citenamefont{Crocker et~al.}(1999)\citenamefont{Crocker, Matteo,
  Dinsmore, and Yodh}}]{Crocker99}
\bibinfo{author}{\bibfnamefont{J.}~\bibnamefont{Crocker}},
  \bibinfo{author}{\bibfnamefont{J.}~\bibnamefont{Matteo}},
  \bibinfo{author}{\bibfnamefont{A.}~\bibnamefont{Dinsmore}}, \bibnamefont{and}
  \bibinfo{author}{\bibfnamefont{A.}~\bibnamefont{Yodh}},
  \bibinfo{journal}{Phys. Rev. Lett.} \textbf{\bibinfo{volume}{82}},
  \bibinfo{pages}{4352} (\bibinfo{year}{1999}).

\bibitem[{\citenamefont{Marconi and Tarazona}(1999)}]{marconi99}
\bibinfo{author}{\bibfnamefont{U.~M.~B.} \bibnamefont{Marconi}}
  \bibnamefont{and} \bibinfo{author}{\bibfnamefont{P.}~\bibnamefont{Tarazona}},
  \bibinfo{journal}{J. Chem. Phys.} \textbf{\bibinfo{volume}{110}},
  \bibinfo{pages}{8032} (\bibinfo{year}{1999}).

\bibitem[{\citenamefont{Marconi and Tarazona}(2000)}]{marconi00}
\bibinfo{author}{\bibfnamefont{U.~M.~B.} \bibnamefont{Marconi}}
  \bibnamefont{and} \bibinfo{author}{\bibfnamefont{P.}~\bibnamefont{Tarazona}},
  \bibinfo{journal}{J. Phys.: Condens. Matter} \textbf{\bibinfo{volume}{12}},
  \bibinfo{pages}{A413} (\bibinfo{year}{2000}).

\bibitem[{\citenamefont{Archer and Rauscher}(2004)}]{archer04b}
\bibinfo{author}{\bibfnamefont{A.~J.} \bibnamefont{Archer}} \bibnamefont{and}
  \bibinfo{author}{\bibfnamefont{M.}~\bibnamefont{Rauscher}},
  \bibinfo{journal}{J. Phys. A: Math. Gen.} \textbf{\bibinfo{volume}{37}},
  \bibinfo{pages}{9325} (\bibinfo{year}{2004}).

\bibitem[{\citenamefont{Rauscher et~al.}(2006)\citenamefont{Rauscher,
  Kr{\"u}ger, Dominguez, and Penna}}]{rauscher06b}
\bibinfo{author}{\bibfnamefont{M.}~\bibnamefont{Rauscher}},
  \bibinfo{author}{\bibfnamefont{M.}~\bibnamefont{Kr{\"u}ger}},
  \bibinfo{author}{\bibfnamefont{A.}~\bibnamefont{Dominguez}},
  \bibnamefont{and} \bibinfo{author}{\bibfnamefont{F.}~\bibnamefont{Penna}}
  (\bibinfo{year}{2006}), \bibinfo{note}{in preparation}.

\bibitem[{\citenamefont{Morse and Feshbach}(1953)}]{morsefeshbach}
\bibinfo{author}{\bibfnamefont{P.~M.} \bibnamefont{Morse}} \bibnamefont{and}
  \bibinfo{author}{\bibfnamefont{H.}~\bibnamefont{Feshbach}},
  \emph{\bibinfo{title}{Methods of theoretical physics}}
  (\bibinfo{publisher}{McGraw-Hill}, \bibinfo{address}{New York},
  \bibinfo{year}{1953}).

\bibitem[{\citenamefont{Aguirre and Murphy}(1973)}]{aguirre73}
\bibinfo{author}{\bibfnamefont{J.~L.} \bibnamefont{Aguirre}} \bibnamefont{and}
  \bibinfo{author}{\bibfnamefont{J.~T.} \bibnamefont{Murphy}},
  \bibinfo{journal}{J. Chem. Phys.} \textbf{\bibinfo{volume}{59}},
  \bibinfo{pages}{1833} (\bibinfo{year}{1973}).

\end{thebibliography}
%\bibliographystyle{apsrev}
\end{document}